\documentstyle[lcws2002]{article}

\RequirePackage{epsfig}
\def\be{\begin{equation}}
\def\ee{\end{equation}}
\def\bea{\begin{eqnarray}}
\def\eea{\end{eqnarray}}
\begin{document}

\title{Questions about the Measurement of the $\rm{e}^{+}\rm{e}^{-}$  Luminosity Spectrum at a future linear Collider.\footnote{Combined 
contribution to the Machine Detector Interface and Calorimeter sessions at LCWS2002.(The Linear Collider Workshop, Jeju Island, Korea, 26-30 August 2002.)}}

\author{S.~T.~Boogert, D.~J.~Miller}
\address{Department of Physics and Astronomy, University College London, Gower Street,\\ London WC1E 6BT, 
England}


\maketitle\abstracts{
Important analyses at a future linear collider, including top-quark and W-boson mass measurements will depend upon the 
precise determination of the luminosity spectrum. This can be done, in principle, in the planned detectors. We 
review the problems to be solved, both beam-related and detector-related.
}

\section{First Order Optimism}
Since the technique was suggested \cite{frarymiller:1991} at the first LCWS,  a number of authors have confirmed 
\cite{kurihara:1993,cinabro:199x,moenig:2000} 
that the measurement of the acollinearity of the final state electron and positron in Bhabha scattering, in 
conjunction with suitable beamline spectrometry, should enable the luminosity spectrum to be extracted 
with sufficient precision to allow the top quark mass to be determined with a precision of a few tens of 
MeV \cite{martinez:2002}.  It may even allow the W boson mass to be measured to around 6 
MeV \cite{tesla:2001:tdr}. Other analyses will require similar precision. But none of these studies has incorporated realistic 
modelling of the fluctuations which may occur in the collisions of the beam bunches, nor have they simulated 
the measurement of the outgoing particles in an actual detector design.  In preparing to do a more detailed 
study we present here some of the issues and effects which may threaten the optimism of the previous studies, 
with the expectation that colleagues will persuade us that some of our doubts are ill-founded, or will draw 
attention to other potential difficulties that we have overlooked. (Inputs after the talks at LCWS are 
included.)     

\section{Sources of Energy Spread}
\begin{figure}[hbt]
\centerline{\epsfxsize=6.5cm \epsfbox{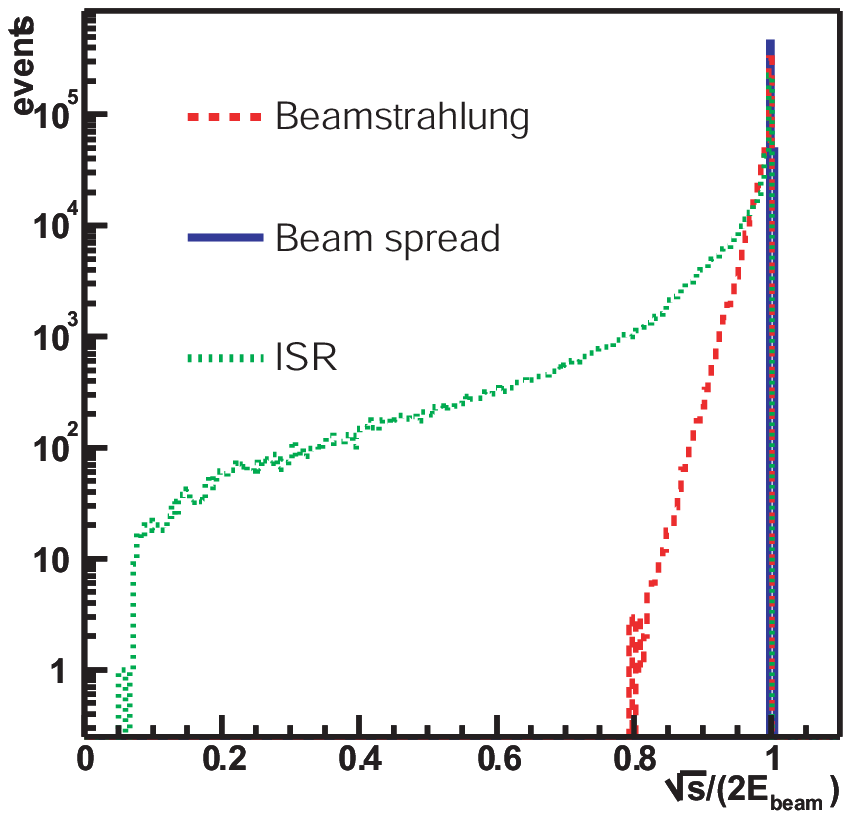} \epsfxsize=6.5cm \epsfbox{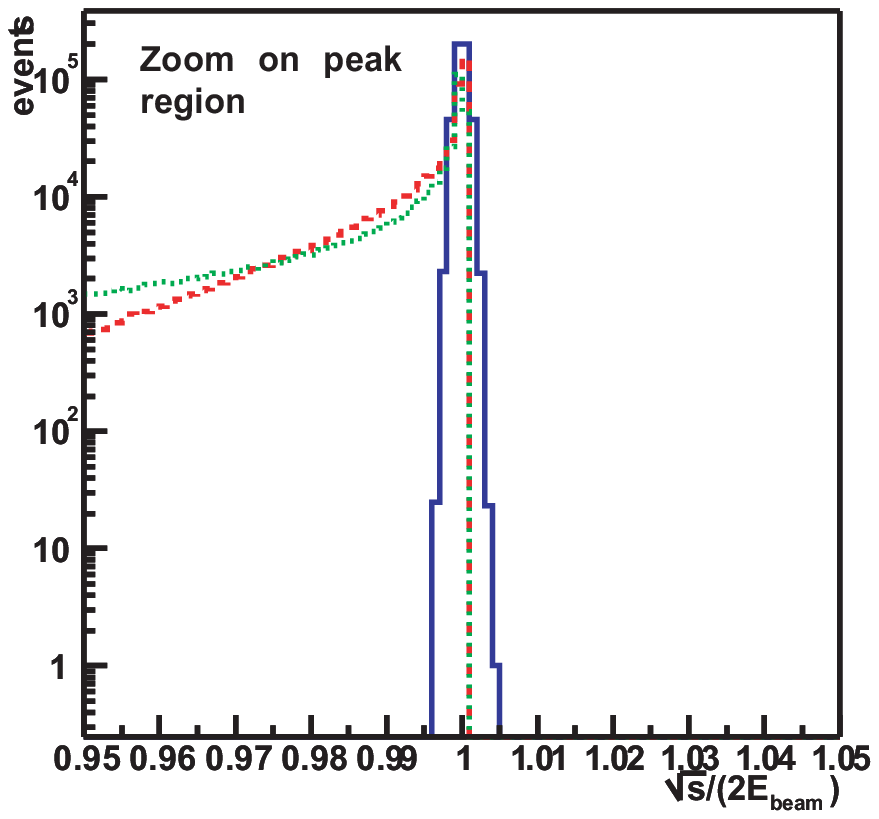}}
\caption{Sources of beam energy loss, for the TESLA design running at 350 GeV.
\label{fig:elosssource}}

\end{figure}

Figure \ref{fig:elosssource} shows an example of the three effects which cause the collision energy at 
a linear collider to be shifted from its nominal value.  Initial State Radiation (ISR) is inescapable, 
and calculable to high precision.  If it were the only effect present there would be no need to measure it. 
Beamstrahlung is essentially synchrotron radiation by individual particles in one bunch due to the high 
electromagnetic fields generated by the charge of the opposing bunch.  In the planned colliders
\cite{tesla:2001:tdr,nlc:2002:rotnlc}, between 30\% and 50\% of collisions have no beamstrahlung, even though the mean 
beamstrahlung loss is 1\% to 3\%, with a long tail of large losses.  This lossless spike is what allows us to contemplate 
precise mass measurements from threshold scans.  The linac itself will introduce a beam energy spread of 
between $\sim0.05$ percent (TESLA \cite{tesla:2001:tdr}) and $\sim 0.3$ percent (X-band \cite{nlc:2002:rotnlc}).  
This is unlikely to be Gaussian and may vary from bunch to bunch.  In TESLA at high energy the $e^{-}$ beamspread is 
increased to $\sim0.15$\% in the helical undulator used to generate polarised positrons (the case shown in Fig. 
\ref{fig:elosssource}), though when running at the WW threshold  this can be avoided.

\section{Measurement of the Absolute Beam Energy} \label{sec:absbe}
 The absolute incoming beam momenta can be measured in upstream spectrometers before 
the collision point, in a similar way to what has been done at LEP \cite{dehning:2000:lepspec}, or in downstream 
spectrometers like that used at Mark II \cite{kent:1990:wisrd}. To use a downstream spectrometer for precise measurements 
it would be necessary to include single electron and positron bunches in the trains without their opposing 
positron and electron bunches, to avoid the disruption, ISR and beamstrahlung from collisions. 
Such precise spectrometers (200 ppm is the goal, which should be enough to get $M_W$ to 6 MeV) are essential to 
measure the absolute incoming energy and its spread. But there is no physics process which has 
good enough energy resolution and sufficiently high statistics to make direct event-based measurements of the 
luminosity spectrum $\partial L/\partial\sqrt{s}$ for individual data-taking runs.

A downstream spectrometer in the spent beam after collision may also be able to monitor the effects of 
beamstrahlung and ISR but it is not clear how this can be done precisely enough for physics analysis. 
Fortunately, as explained below, the acollinearity of Bhabha scattering in the endcap region of the 
detector gives a high statistics event-based technique which is sensitive to the relative beamspread from 
all three sources.
This note is not primarily concerned with the spectrometer measurements of absolute energy and spread, 
but we here list a number of crucial questions about them which are being addressed elsewhere \cite{torrencehertzbach:pc}:
\begin{itemize}
\item  Can we get both mean energy and beamspread shape on a bunch-by-bunch basis or just train-by-train?
\item Since all the bunches in a train do not collide optimally, how do we calculate luminosity-weighted 
energies and spreads for each physics run?  This may be posible with input from a high-rate 
small angle Bhabha detector like the TESLA LCAL \cite{tesla:2001:tdr}. 
\item How is TESLA different from X-band?  The zero crossing angle, followed by an extraction 
kicker, makes a downstream spectrometer much less effective, but the hundreds of nanoseconds 
between separate bunches may allow an upstream spectrometer to resolve them in a way which cannot 
be done for X-band, and allow the LCAL to luminosity-weight each bunch.
\item  How much will the incoming energy wander from train to train or from bunch to bunch?
\item  How much will the nongaussian beamspread shape vary between bunches and between trains?
\end{itemize}

\section{Basic Principles of the Acollinearity Method} \label{sec:acol}
The acollinearity angle $\theta_{A}$ is defined in Figure 2.  For  $\theta_{A} << \theta$ we have $ \theta_{A} 
= (\Delta p/p_b) \sin \theta$, where $\Delta p = p_{+} - p_{-}$, i.e. the missmatch between the two beam 
momenta at collision.  The quantity needed for physics is $\sqrt{s} \simeq p_+ + p_-$.  For small $ 
\theta_A$ and Gaussian errors, $\sigma_{\sqrt{s}} \simeq \sigma_{\Delta p} \simeq \sigma_{ \theta_A} 
{p_b}/{\sin \theta} \simeq \sqrt{2} \sigma_{p_b}$.  So with a given angular resolution $\sigma_{ 
\theta_A}$ the error on $\sqrt{s}$ blows up at small values of the scattering angle $\theta$.  This means 
that the best sensitivity to $\sqrt{s}$ will be in the endcap region ($100 \leq \theta \leq 450$ 
milliradians) of the detector. In this region there is a large t-channel contribution to Bhabha scattering, 
giving a rate about 400 times the pointlike s-channel rate which governs $\mu^{+}\mu^{-}$, $t\overline{t}$ and many of the 
other interesting SM or SUSY processes. As discussed in \ref{sec:cal} 
below, the detector angular resolutions in this region are expected to be as good as is needed not only for 
the top mass measurement but also for $M_W$.
\begin{figure}[hbt]
\centerline{\epsfxsize=5.5cm \epsfbox{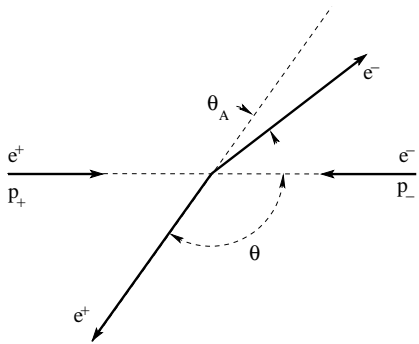}}
\caption{Definition of acolinearity angle $\theta_{A}$
\label{fig:acoldef}}

\end{figure}

\begin{figure}[hbt]
\centerline{\epsfxsize=6.5cm \epsfbox{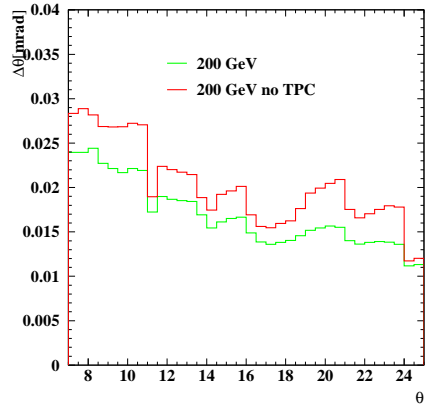}}
\caption{Angular resolution of the forward tracking detectors at TESLA
\label{fig:ftrackres}}

\end{figure}

\section{Beam-related Questions about the Acollinearity Method} \label{sec:beam}
\begin{figure}[hbt]
\centerline{\epsfxsize=6.5cm \epsfbox{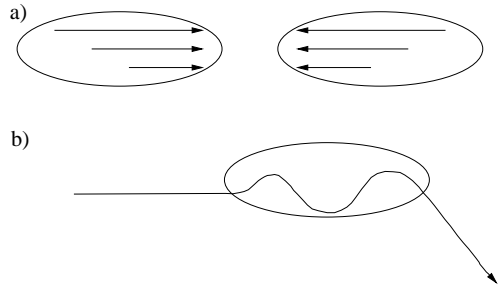}}
\caption{How the effects of dispersion could be removed by disruption
\label{fig:beffects}}

\end{figure}

Positive correlations between the event-to-event shifts in the momenta $p_+$ and $p_-$ can give rise 
to reduced values of $\Delta p$ and of the acollinearity, even when the shifts in $\sqrt{s}$ are large. 
At least three sources of such correlations have been suggested.  
\begin{itemize}
\item Dispersion effects.  If there were dispersion at the final focus then each beam might 
(see Fig. \ref{fig:beffects}a) have slightly higher momenta on one edge of the bunch and slightly lower 
momenta on the other edge \cite{telnov:pc}.  But disruption, as sketched (see Fig. \ref{fig:beffects}b), actually causes 
particles from one beam to oscillate violently as they pass through the opposing bunch, and this may obliterate 
the effect of dispersion.
\item Early-late Correlation. Evidence has already been reported \cite{moenig:2000} for a correlation in simulated 
events between the collision energies of $e^+$ and $e^-$ which meet early in the crossing of two bunches and 
those which meet late in the crossing, when particles in both bunches are more likely to have had beamstrahlung 
losses.  Coarse information from small angle detectors on the bunch-by-bunch luminosity may be useful in 
correcting for this effect.
\item  Bananas.  If beams become misaligned in the LINAC then the bunches may develop `banana-shaped' tails which are expected 
to have slightly different energy from the main part of the bunch.
\end{itemize}

It may also be possible (see discussion of angular precision below) to observe an azimuthal variation in the 
spread of acollinearity due to the much greater disruption in the horizontal plane compared with the vertical 
\cite{schulte:1997:thesis}. All of these effects will be studied with the full 
Guinea Pig \cite{schulte:1997:thesis} beam-beam interaction simulation program.  

\section{Calorimeter-related Questions about the Acollinearity Method}\label{sec:cal}
\begin{figure}[hbt]
\centerline{\epsfxsize=8.0cm \epsfbox{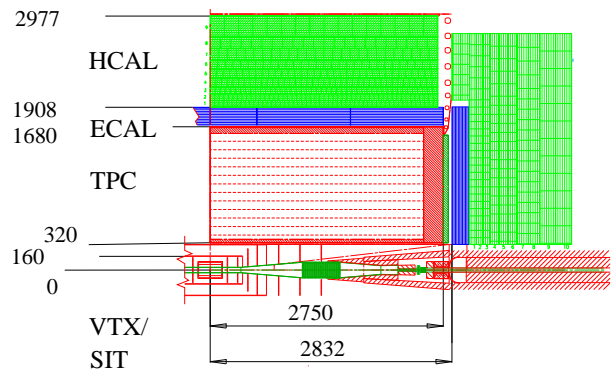}}
\caption{A Quadrant of the TESLA detector
\label{fig:tesladet}}

\end{figure}

The first question is; which detectors to use to measure the luminosity spectrum?  
As mentioned in \ref{sec:acol} above, the error on $\sqrt{s}$ for a given angular error on the outgoing 
tracks is proportional to $1/\sin \theta$. In the TESLA detector (Fig. \ref{fig:tesladet}), the best available angular 
resolution ($\sim 20-30 \hspace{0.1cm}\mu\rm{radian}$) will come from the Forward Tracking discs and the 
TPC central tracker, so the acollinearity method will give the best results in this angular region.
If this precision can be exploited \cite{moenig:2000} there would be no trouble 
in measuring the top quark mass to the desired precision, and the W mass should not be too difficult.  
But a luminosity measurement requires high efficiency, and the track finding efficiency is unlikely to be 
much above 90\%, especially in the forward region where background tracks spiral around field lines and give 
extra hits.  Fortunately the whole of the region is backed by the CALICE electromagnetic calorimeter
\cite{calice:2001:proposal} with fine granularity, both laterally and longitudinally.  
This should have 100\% efficiency for high energy electrons, so it can be used to map the efficiency of the tracker.  

In addition, CALICE will have an angular resolution for high energy electrons of about 0.5 milliradians, so it 
might by itself provide an adequate acollinearity measurement for the top quark mass measurement, if it can be 
surveyed to sufficient precision.  Experience from OPAL \cite{attree:1994:wet} shows that a precision tracker in front of a 
calorimeter can provide such a survey.  We are beginning a study of the performance of CALICE in this region, 
with special attention to the degradation of the angle measurement in CALICE which will come from scattering 
and delta-ray production in the field cage and end plates of the TPC.  This produces soft electrons which run 
along magnetic field lines to hit the calorimeter at lower radius than the main shower.

The LAT calorimeter at TESLA covers the range $30<\theta <87$ milliradians, similar to the coverage of the LEP 
and SLC luminosity monitors.  Like them it will have a high enough Bhabha scattering rate to exceed interesting 
physics channels even at the peak of the $Z^{0}$ resonance.  It is therefore an important absolute luminosity 
monitor.  Its angular resolution will be similar to CALICE and it will have less dead material in front of it.  
But this 
angular region is not so good for measuring the luminosity spectrum by the acollinearity method because of the 
$1/\sin \theta$ factor mentioned above.  It has pointed out \cite{blair:pc} that the summed energies 
of the two showers in LCAL Bhabha events will give a measure of $\sqrt{s}$ with a precision of 1-2\%.  Although 
this is not precise enough to resolve the spike in the luminosity spectrum (Fig 1) which is needed for 
measurements of the top mass or the W mass, it is sufficient to constrain a large part of the ISR and beamstrahlung 
tail.

The TESLA design also includes a very small angle calorimeter, the LCAL, spanning 5 to 27 milliradians.  Even 
with a 4T magnetic field this detector will be hit on every beam crossing by an intense background flux of soft 
photons and electrons from the beamstrahlung and pair production processes.  This will make it a poor device for 
absolute luminosity measurement because the efficiency for the Bhabha signal will be badly measured, especially 
close to the lower angle cutoff.  But its background rate may be high enough to give a rough bunch-by-bunch 
measure of the luminosity which can be used to weight the spectrometer output (see \ref{sec:absbe} above).  
It may also be useful in correcting for the early-late correlation (see \ref{sec:beam} above). 

\section{Unfolding Variables}
It has been suggested \cite{blair:pc}, that the best approach to the determination of 
$\partial L/\partial \sqrt{s}$ for any event sample will be to unfold the spectrum from measurements 
of the distributions of all available sensitive variables.  These would include: a) the sums of Bhabha shower 
energies from LAT, LCAL and CALICE (to constrain the ISR + beamstrahlung tail); b) results from the beam spectrometers 
on both absolute energy and spread, with luminosity weighting; c) an appropriate variable derived 
from the acollinearity of Bhabha scatters in the endcap region.  Two such variables have been discussed:  
$\Delta p$, as defined in \ref{sec:acol} above \cite{frarymiller:1991}, or $s^{\prime}$, the invariant mass of the final $
e^+ e^-$ on the assumption that a single radiated gamma has been lost along the beam direction \cite{moenig:2000}.

\vspace{-0.115cm}
\section{Plans}
Guinea Pig \cite{schulte:1997:thesis} and other simulations of the collision process are being set up to produce samples which contain 
all of the possible beam-based effects that could bias measurements of the luminosity spectrum.  These events 
will be passed through a realistic model of the forward regions of TESLA detector, including CALICE, and 
reconstructed to extract the relevant variables.  The reconstructed or unfolded luminosity spectrum will be 
compared with the Monte Carlo truth.  The consequences for physics of any disagreement between truth and 
measurement will be investigated, especially for the top and W mass measurements.

\end{document}